\documentstyle[12pt,bezier]{article}
\def\beq{\begin{equation}}
\def\eeq{\end{equation}}

\def\ap#1#2#3 {Ann. Phys. (NY) {\bf#1} (19#2) #3}
\def\apj#1#2#3 {Astrophys. J. {\bf#1} (19#2) #3}
\def\apjl#1#2#3 {Astrophys. J. Lett. {\bf#1} (19#2) #3}
\def\app#1#2#3 {Acta. Phys. Pol. {\bf#1} (19#2) #3}
\def\ar#1#2#3 {Ann. Rev. Nucl. Part. Sci. {\bf#1} (19#2) #3}
\def\cpc#1#2#3 {Computer Phys. Comm. {\bf#1} (19#2) #3}
\def\err#1#2#3 {{\it Erratum} {\bf#1} (19#2) #3}
\def\ib#1#2#3 {{\it ibid.} {\bf#1} (19#2) #3}
\def\jmp#1#2#3 {J. Math. Phys. {\bf#1} (19#2) #3}
\def\ijmp#1#2#3 {Int. J. Mod. Phys. {\bf#1} (19#2) #3}
\def\jetp#1#2#3 {JETP Lett. {\bf#1} (19#2) #3}
\def\jpg#1#2#3 {J. Phys. G. {\bf#1} (19#2) #3}
\def\mpl#1#2#3 {Mod. Phys. Lett. {\bf#1} (19#2) #3}
\def\nat#1#2#3 {Nature (London) {\bf#1} (19#2) #3}
\def\nc#1#2#3 {Nuovo Cim. {\bf#1} (19#2) #3}
\def\nim#1#2#3 {Nucl. Instr. Meth. {\bf#1} (19#2) #3}
\def\np#1#2#3 {Nucl. Phys. {\bf#1} (19#2) #3}
\def\pcps#1#2#3 {Proc. Cam. Phil. Soc. {\bf#1} (#2) #3}
\def\pl#1#2#3 {Phys. Lett. {\bf#1} (19#2) #3}
\def\prep#1#2#3 {Phys. Rep. {\bf#1} (19#2) #3}
\def\prev#1#2#3 {Phys. Rev. {\bf#1} (19#2) #3}
\def\prl#1#2#3 {Phys. Rev. Lett. {\bf#1} (19#2) #3}
\def\prs#1#2#3 {Proc. Roy. Soc. {\bf#1} (19#2) #3}
\def\ptp#1#2#3 {Prog. Th. Phys. {\bf#1} (19#2) #3}
\def\ps#1#2#3 {Physica Scripta {\bf#1} (19#2) #3}
\def\rmp#1#2#3 {Rev. Mod. Phys. {\bf#1} (19#2) #3}
\def\rpp#1#2#3 {Rep. Prog. Phys. {\bf#1} (19#2) #3}
\def\sjnp#1#2#3 {Sov. J. Nucl. Phys. {\bf#1} (19#2) #3}
\def\spj#1#2#3 {Sov. Phys. JEPT {\bf#1} (19#2) #3}
\def\spu#1#2#3 {Sov. Phys. Usp. {\bf#1} (19#2) #3}
\def\zp#1#2#3 {Zeit. Phys. {\bf#1} (19#2) #3} 

\begin{document}
\begin{titlepage}
\begin{center}
{\Large \bf Theoretical Physics Institute \\
University of Minnesota \\}  \end{center}
\vspace{0.3in}
\begin{flushright}
TPI-MINN-96/30-T \\
UMN-TH-1526-96 \\
December 1996
\end{flushright}
\vspace{0.4in}
\begin{center}
{\Large \bf  $O(m_c^{-2})$ nonperturbative corrections to the inclusive  
rate of the decay $B \to X_s \, \gamma$.\\}
\vspace{0.2in} 
{\bf M.B. Voloshin  \\ } 
Theoretical Physics Institute, University of Minnesota, Minneapolis, MN 
55455 \\ and \\ 
Institute of Theoretical and Experimental Physics, Moscow, 117259 \\[0.2in]

%{\bf   Abstract  \\ }
\end{center}

\begin{abstract}
It is shown that the inclusive rate of the rare weak radiative decays $B \to 
X_s \, \gamma$ contains a series of nonperturbative corrections, 
whose `short distance' scale is set by $m_c^{-1}$, rather than by $m_b^{- 1}$. 
The first correction in this series is expressed through the chromomagnetic 
interaction of the $b$ quark inside the $B$ meson and the relative magnitude 
of the effect is determined by the ratio $\langle {\overline B} | \bar b \, 
\sigma \cdot G \, b | B \rangle / m_c^2$. Though the magnitude of this first 
correction is suppressed by a numerical coefficient, the sensitivity of the 
decay rate to the distance scale $m_c^{-1}$ may significantly limit the accuracy 
of purely perturbative predictions for the rate.
\end{abstract} 
\end{titlepage}

The rare radiative decays $B \to X_s \, \gamma$ associated with the 
underlying FCNC quark process $b \to s \, \gamma$ are well known to be 
sensitive to the top quark mass as well as to a possible `new physics' that 
might exist beyond the Standard Model. The current state of the experimental 
knowledge about these processes and of their understanding within the 
Standard Model can be summarized by the measured value$^{\cite{cleo}}$ of 
the inclusive branching ratio of these decays
\beq
B(B \to X_s \, \gamma) = (2.32 \pm 0.57 \pm 0.35) \times 10^{-4}
\label{exp}
\eeq
and by the perturbative Standard Model quark decay rate, whose 
calculation has been recently taken$^{\cite{cmm}}$ to the next-to-leading 
order (NLO) in $\alpha_s \, \ln (m_W/m_b)$,
\beq
B(b \to X_s \, \gamma) = (3.28 \pm 0.33) \times 10^{-4}~~.
\label{theo}
\eeq
Additionally, it has been argued$^{\cite{fls}}$ that the nonperturbative 
effects, which distinguish between the heavy quark decay and the decay of 
a heavy meson, are of relative magnitude $\Lambda_{QCD}^2/m_b^2$. Thus in
the subsequent studies (see Ref.\cite{cmm} and references therein) it
has been assumed that the perturbative total quark decay rate approximates the
total inclusive meson decay rate with accuracy better than 10\%, though the
$O(1/m_b^2)$ nonperturbative effects are naturally expected to be
significant in the photon spectrum near the endpoint$^{\cite{dsu}}$.

The purpose of this paper is to point out that there in fact exists a 
series of nonperturbative corrections to the inclusive rate of the decays 
$B \to X_s \, \gamma$, whose `short distance' scale is set by
$m_c^{-1}$, rather than by $m_b^{-1}$. This series thus generates an
expansion for the corrections to the decay rate in
powers of $\Lambda_{QCD}/m_c$ and of $\Lambda_{QCD} \, m_b/m_c^2$. 
The first term in this expansion is
$O(\Lambda_{QCD}^2/m_c^2)$ and is found in terms of the known chromomagnetic 
energy of the $b$ quark. The subsequent terms, however are expressed
trough unknown matrix elements of operators of higher dimension. 
The relative magnitude of the calculable first correction
to the total decay rate is found here in the lowest 
order in $\alpha_s$ and is given by
\beq
{\delta \Gamma(B \to X_s \, \gamma) \over \Gamma(b \to X_s \, \gamma)} =
{1 \over 27 \, C_7}\, { \mu_g^2 \over m_c^2} \approx -0.025 ~~,
\label{result}
\eeq
where $\mu_g^2$ is the standard parameter$\cite{buv}$ for the strength of the 
chromomagnetic interaction of the $b$ quark inside the hadron: 
$\mu_g^2 \equiv \langle  B |
\bar b \, g\, \sigma_{\mu \nu} G_{\mu \nu}^a \, 
(\lambda^a/2) \, b | B \rangle/2 =
3 \, (M_{B^*}^2-M_{B}^2)/4 \approx 0.4\, GeV^2$. (The nonrelativistic
normalization for the heavy quark states is assumed throughout this paper, 
so that $\langle B | b^\dagger b | B \rangle = 1$.)
The quantity $C_7$ in
the denominator in eq.(\ref{result}) is the coefficient of the operator
\beq
P_7={e \over 16\pi^2}\, m_b \, (\bar s_L \, \sigma_{\mu \nu} \, b_R) \,
F_{\mu \nu}
\label{p7}
\eeq
in the perturbative effective Lagrangian for the $b$ quark
decay normalized at a scale $\mu \sim m_b$:
\beq
L_{eff} =  {4\, G_F \over \sqrt{2}} \, V_{ts}^* \, V_{tb} \sum_{i=1}^8
C_i (\mu) \, P_i (\mu)~~,
\label{leff}
\eeq
and the values $C_7 = -0.30\,^{\cite{cmm}}$ and $m_c \approx 1.4 \,
GeV$ are used in the numerical estimate in eq.(\ref{result})\footnote{The 
notation and conventions of Ref.\cite{cmm} are used here as well as 
the numerical value  of $C_7(m_b)$. Furthermore, eq.(\ref{result}) 
is written in the approximation where $\Gamma(b \to X_s \, \gamma) 
= \Gamma(b \to s \, \gamma)$ 
and the latter rate is generated only by the operator $P_7$. 
In NLO the rate also receives a small contribution from other operators 
in $L_{eff}$. However, since the correction is calculated here only 
in the leading order in $\alpha_s$, taking into account those 
additional contributions to $\Gamma(b \to X_s \, \gamma)$  
would exceed the accuracy of the present calculation. }.
Clearly, the results of the present paper, in particular
eq.(\ref{result}), are at variance with the generally accepted viewpoint 
that the relative magnitude of the
nonperturbative corrections to the inclusive decay rate of $B \to X_s \,
\gamma$ goes to zero in the limit $m_b \to \infty$. The latter behavior
would be true if the relevant contributions to the $B$ meson decay were
exhausted by the effective Lagrangian in eq.(\ref{leff}), which is the
case for the perturbative $b$ quark decay. At the level of
nonperturbative effects there are additional terms in the
effective Lagrangian, among which there are the ones of the type
considered in this paper, whose distance scale is set by $m_c^{-1}$, so
that their (relative) contribution does not vanish in the limit $m_b \to
\infty$. The correction in eq.(\ref{result}) is small due to a small
numerical factor. Apriori there is no reason to expect that similar
coefficients are small in subsequent terms. Moreover, the subsequent
terms contain average values of operators of the generic form $(\bar b
(q \, D)^n \, G b)/m_c^{2n+2}$, 
where $q$ is the momentum of the gluon and the covariant
derivative $D$ is acting on the gluon field tensor $G$. After averaging
over the phase space in the decay these contributions produce series in
$(\Lambda_{QCD}^2/m_c^2)\, (\Lambda_{QCD} \, m_b/m_c^2)^n$. Since in
reality the parameter $\Lambda_{QCD} \, m_b/m_c^2$ is of order one,
these terms bring a considerable uncertainty in predictions for the
decay rate.

The consideration of the nonperturbative effects in the inclusive decay
rate is performed using the standard method$^{\cite{sv0,vs}}$. 
Namely, both $m_b$ and $m_c$ are
assumed to be large: $m_b,\, m_c \gg \Lambda_{QCD}$, and one performs
the operator product expansion in the inverse powers of the heavy quark
masses for the effective operator
\beq
T=2 \,{\rm Im} \, \left [ i \int d^4x \, e^{iqx} \, T \left \{ 
L_{eff}^\dagger(x), L_{eff}(0) \right \} \right ]~.
\label{teff}
\eeq
The total decay rate for a heavy hadron $X_b$ is then given by the
diagonal matrix element of $T$ over $X_b$ :
\beq
\Gamma(X_b) = \langle X_b | T | X_b \rangle ~.
\label{gam}
\eeq
Different types of decay are separated within this method by picking up
the appropriate terms in $L_{eff}$ to correlate in eq.(\ref{teff}). 
The leading term in the expansion
generates the perturbative `parton' decay rate, while the terms suppressed
by inverse powers of heavy quark masses describe the nonperturbative
corrections to the decay rate.

In the leading logarithm approximation (LLA) the perturbative quark 
decay rate of $B \to X_s \, \gamma$ is obtained by retaining in
$L_{eff}$ only the term with the operator $P_7$ in eq.(\ref{leff}) and
substituting it in the correlator $T$ in eq.(\ref{teff}) with the result
\beq
T^{(0)}_{s\gamma}= {\alpha \over 32 \, \pi^4} \, G_F^2 \, m_b^5 \, 
|V_{ts}^* \,V_{tb}|^2 \, |C_7|^2 \, (\bar b \, b)~.
\label{t0}
\eeq
This expression gives the standard perturbation theory formula
for the inclusive rate of $B \to X_s \, \gamma$ upon using noticing 
that up to terms $O(m_b^{-2})$ one has 
$\langle B |(\bar b \, b)| B \rangle =1 \,^{\cite{buv}}$.

The reasoning here for terms suppressed by only $m_c^{-2}$ rather than
by $m_b^{-2}$ is based on considering the `gluon - photon penguin' type
of mechanism, which provides additional to eq.(\ref{leff}) terms in the
effective Lagrangian. This mechanism arises through graphs of the type
shown in Fig. 1. The gluon momentum $k$ in this graph is assumed to be
small, and the expansion in this momentum gives rise to a series of 
operators in $L_{eff}$ with the gluon field
strength tensor $G$, and its derivatives. One can readily see, 
that already starting from first such term the loop integration 
for the coefficients in this
expansion is convergent in the local four-fermion 
limit of the weak interaction. In particular, the corresponding
graph with the top quark becomes irrelevant. In the local limit of the
weak interaction the loop with the $c$ quark is kinematically equivalent 
(after the Fierz
transform of the four-fermion vertex) to the $c$ quark loop for coupling
of axial current to two vector currents. Since the photon momentum $q$ is
on-shell: $q^2=0$, and the gluon momentum is infinitesimal, the external
kinematical invariants for the loop, $q^2$, $k^2$, and $(q+k)^2$ are either
zero or infinitesimal. Thus the coefficients of expansion of the loop in
powers of $k$ are determined by the mass of the quark in the loop, i.e.
by $m_c$. Furthermore, an expansion of the loop in the invariant $(q+k)^2$ 
gives rise to terms with powers of the ratio $(q \cdot k)/m_c^2$,
resulting in the operator terms of the form $(\bar b \,(q \, D)^n
\, G \, b)/m_c^{2n+2}$, which give rise to corrections to the rate of
the relative magnitude $(\Lambda_{QCD}^2/m_c^2)\, (\Lambda_{QCD} \,
m_b/m_c^2)^n$.

By calculating the graph of Fig. 1 and the one with the photon 
and the gluon permuted, and expanding to the first order in the 
gluon momentum $k$, one explicitly finds the first term 
in this series of corrections 
as an additional contribution to $L_{eff}$ of the form
\beq
L_{eff}^{(s g \gamma)}= {e \, Q_c \over 16 \, \pi^2} \,\sqrt{2} \, G_F \, 
V_{cs}^* \,V_{cb} \, (\bar s_L \, \gamma_\mu \, {\lambda^a \over 2} \,
b_L) \, {1 \over 3 m_c^2} \, g \,G_{\nu \lambda}^a \, \epsilon_{\mu \nu \rho
\sigma} \,i \partial_\lambda F_{\rho \sigma}~~,
\label{lsgg}
\eeq
where $Q_c=2/3$ is the electric charge of the charmed quark and the
expression is written under the convention about the sign of $G_F$ as in
Ref.{\cite{cmm}}, i.e. where the bare four-fermion Lagrangian is written
as $L(b \to c \bar c s)=-(4 G_F/\sqrt{2})\, V_{cs}^* \,V_{cb} \, 
(\bar s_L \gamma_\mu c_L)\, (\bar c_L \gamma_\mu b_L)$.

When correlated according to eq.(\ref{teff}) through the on-shell 
(massless) $s$ quark and the photon with the term with the operator
$P_7$ in $L_{eff}$ of eq.(\ref{leff}) the newly found term $L_{eff}^{(s
g \gamma)}$ gives an interference contribution to the effective operator
$T$. This contribution is found here as
\beq
T^{(1)}_{s\gamma}=-{\alpha \over 32 \, \pi^4} \, G_F^2 \, m_b^5 \,{1 \over 27}
{\rm Re} \, \left [ V_{cs}^* \,V_{cb} \, V_{ts} \, V_{tb}^* \, C_7
\right ] \, {{( \bar b \, g\, \sigma_{\mu \nu} G_{\mu \nu}^a \, 
{\lambda^a \over 2} \, b)} \over 2 \, m_c^2}~.
\label{dgam}
\eeq
Ignoring the terms in the unitarity relation 
for quark mixing associated with the $u
\bar u$ pair contribution, which are of the relative order
$(|V_{ub}/V_{cb}|) \, \sin \theta_c $, one can write 
$V_{cs}^* \,V_{cb} = - V_{ts}^* \, V_{tb}$. Then a comparison of the
equations (\ref{dgam}) and (\ref{t0}) gives the expression in
eq.(\ref{result}) for the relative nonperturbative correction.

Proceeding to a discussion, we first remark that
the ignored contribution of the $u \bar u$ pair in fact leads to a
hopefully small, but presently uncalculable effect in the rate of $B \to X_s\,
\gamma$. The smallness is obviously related to the suppression in the
mixing, while the uncertainty of the contribution arises, when one
considers the same loop as in Fig. 1, but with the $u$ quark instead of the
charmed one. In no approximation can the mass of the $u$ quark be
considered as large. Thus this mechanism gives rise to an essentially
long distance contribution to the inclusive rate\footnote{Possible 
long distance effects were in fact discussed within a different technique 
in the literature$^{\cite{dht}}$. I believe that the used here systematic 
method  of OPE provides a more reliable approach to calculable terms and
leaves undetermined only a small contribution due to the light $u \bar u$ 
pair, which is strongly suppressed by the relative factor
$(|V_{ub}/V_{cb}|) \, \sin \theta_c \approx 0.02$.}.

Returning to the discussion of the corrections arising from distances
$O(m_c^{-1})$, it can be noticed that in general the effect of such distances 
can also significantly affect the spectrum of inclusive photons through
enlarged distance scale for the amplitude of the photon emission. However
any detailed consideration of this effect is beyond the scope of the
present paper.

The particular numerical value in eq.(\ref{result}) for the 
relative correction to the inclusive rate is most certain to be modified
by the subsequent terms with derivatives of the gluonic tensor. 
Apriori those terms have no parametric smallness in comparison with the
expression in eq.(\ref{result}), since the additional factors in those
terms are given by powers of the parameter $\Lambda_{QCD} \,
m_b/m_c^2 \sim O(1)$. The principal difficulty of evaluating those
subsequent terms is in that we have no knowledge of the matrix elements
of operators with derivatives of the gluonic tensor over the $B$ meson.
Therefore this series of the corrections puts an
essential limit on the present calculability of the decay rate.

It can be noted that the presence of a contribution in the rate due 
to the chromomagnetic
interaction of the $b$ quark can in principle be verified experimentally
by measuring also analogous inclusive rate for the $\Lambda_b$ baryon:
$\Gamma (\Lambda_b \to X_s \, \gamma)$. In $\Lambda_b$ the spin of the
light component of the baryon is zero, thus the average of the
chromomagnetic operator vanishes. Therefore for $\Lambda_b$ the discussed
correction is absent, and the difference between the inclusive decay rates   
$\Gamma (B \to X_s \, \gamma) - \Gamma (\Lambda_b \to X_s \, \gamma)$ is
dominated by the spin-dependent terms in the discussed corrections 
with further small corrections
being $O(m_b^{-2})$ and analogous to those found$^{\cite{buv}}$ 
for the dominant decays. Thus this difference in rates can serve as a
measure of importance of the nonperturbative corrections coming from
distances $O(m_c^{-1})$.

In summary. It has been shown that there exists a set of nonperturbative
corrections to the inclusive rate of the decays $B \to X_s \, \gamma$,
which are determined by the distance scale $O(m_c^{-1})$. The relative
magnitude of these corrections is not vanishing in the limit $m_b \to
\infty$, provided that $m_c$ is kept fixed. The first and calculable, 
although not necessarily dominant, among these
corrections is of the order $O(m_c^{-2})$ and is given in the lowest order
in $\alpha_s$ by eq.(\ref{result}).  The presence of
the distance scale $m_c^{-1}$ in the `penguin' type decays of the $b$
hadrons may have further implication for the photon radiative decays as
well as for the nonleptonic decays of these hadrons.

I am thankful to Mikhail Shifman for useful discussions and to Mark Wise for
pointing out a numerical error in an earlier version of this paper. 
This work is 
supported, in part, by the DOE grant DE-AC02-83ER40105.

{\large \bf Figure caption.} \\
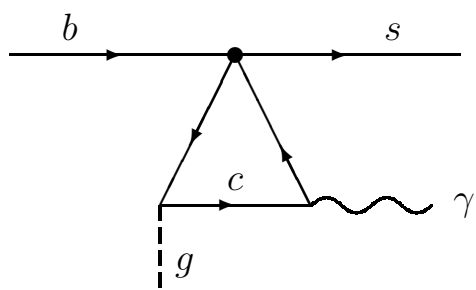
{\bf Figure 1.} A graph for the `gluon-photon penguin'. The heavy dot
represents the four-fermion weak interaction.
At small momentum $k$ of
the gluon the scale for the expansion in powers of $k$ is set by $m_c$.

\newpage
\unitlength=1.00mm
\thicklines
\begin{picture}(100.00,122.00)
\put(40.00,80.00){\line(1,0){60.00}}
\put(70.00,80.00){\line(-1,-2){10.00}}
\put(60.00,60.00){\line(1,0){20.00}}
\put(80.00,60.00){\line(-1,2){10.00}}
\put(48.00,80.00){\vector(1,0){7.00}}
\put(78.00,80.00){\vector(1,0){7.00}}
\put(70.00,80.00){\vector(-1,-2){6.00}}
\put(80.00,60.00){\vector(-1,2){4.00}}
\put(60.00,60.00){\vector(1,0){10.00}}
\put(60.00,60.00){\line(0,-1){2.00}}
\put(60.00,57.00){\line(0,-1){2.00}}
\put(60.00,54.00){\line(0,-1){2.00}}
\put(60.00,51.00){\line(0,-1){2.00}}
\bezier{24}(80.00,60.00)(82.00,62.00)(84.00,60.00)
\bezier{24}(84.00,60.00)(86.00,58.00)(88.00,60.00)
\bezier{24}(88.00,60.00)(90.00,62.00)(92.00,60.00)
\bezier{24}(92.00,60.00)(94.00,58.00)(96.00,60.00)
\put(48.00,82.00){\makebox(0,0)[cb]{{\large $b$}}}
\put(91.00,82.00){\makebox(0,0)[cb]{{\large $s$}}}
\put(70.00,62.00){\makebox(0,0)[cb]{{\large $c$}}}
\put(99.00,60.00){\makebox(0,0)[lc]{{\large $\gamma$}}}
\put(62.00,52.00){\makebox(0,0)[lc]{{\large $g$}}}
\put(70.00,80.00){\circle*{2.00}}
\end{picture}
\vspace{0.5 in}

\centerline{\large \bf Figure 1}

\end{document}